\documentclass[
aps,%
12pt,%
final,%
notitlepage,%
oneside,%
onecolumn,%
nobibnotes,%
nofootinbib,%
superscriptaddress,%
noshowpacs,%
centertags]%
{revtex4}

\usepackage{url}
\usepackage{amstext}

\def\deg{\hbox{$^\circ$}}                  
\def\arcm{\hbox{$^\prime$}}                
\def\farcm{\hbox{$.\mkern-4mu^\prime$}}    
\def\arcs{\hbox{$^{\prime\prime}$}}        

\begin{document}
\selectlanguage{english}

\title{IDENTIFICATION OF NEW CATACLYSMIC VARIABLES\\
IN THE 1RXS AND USNO-B1.0 CATALOGS}

\author{\firstname{D.~V.}~\surname{Denisenko}}
\email{denis@hea.iki.rssi.ru}
\affiliation{%
Space Research Institute of Russian Academy of Sciences, Moscow, Russia
}%
\author{\firstname{K.~V.}~\surname{Sokolovsky}}
\email{ksokolov@mpifr-bonn.mpg.de}
\affiliation{%
Max Planck Institute for Radio Astronomy, Bonn, Germany
}%
\affiliation{%
Astro-Space Center, Lebedev Physical Institute of Russian Academy of Sciences, Moscow, Russia
}%

\begin{abstract}
As a result of applying the original optical variability search method 
on publicly available data, we have found eight new
cataclysmic variables and two possible Optically Violent Variable quasars
among the previously unidentified X-ray sources in the {\em ROSAT} catalog. We
describe the search method and present the characteristics of the newly identified
variable stars. The obtained results demonstrate the large potential of
the concept of Virtual Observatory for identifying new objects of
astrophysical interest.
\end{abstract}

\maketitle

\section{Introduction}

Despite the growing number of cataclysmic variable stars discovered both at
optical wavelengths from the ground (see, for example, Wils et al., 2009,
Drake et al., 2009) and in X-rays from space (Motch et al., 1996, Verbunt et
al., 1997 and a number of other works), many of these objects,
worthy of attention and further study, still remain undiscovered while the
information about their unusual properties can be obtained from the existing
data of various catalogs and archival observations. Following the concept
of Virtual Observatory, we use simple search tools on the archival data
which are publicly available for years to identify new objects of
astrophysical interest.

Since 2005 one of the authors (D.D.) is carrying out a systematic work to
identify the poorly studied X-ray sources from 1RXS catalog which is based
on the data from {\em ROSAT}
(Voges et al., 1999). As a result of this work several dozen new
variable stars have been discovered, including high-amplitude RS~CVn stars
(Denisenko, 2008), Algol type eclipsing binaries, as well as a few
cataclysmic variables -- deeply eclipsing polar 1RXS~J020929.0$+$283243
(Denisenko et al., 2006), dwarf novae 1RXS~J180834.7$+$101041 (Denisenko et
al., 2008),  1RXS~J173006.4$+$033813 (Denisenko et al., 2009),
1RXS~J231935.0$+$364705 (Denisenko, 2009) and several others 
to be published. All these cataclysmic variables were identified in the
archival images of DSS (Digitized Palomar Sky Survey photographic plates)
with the occasional addition of data from the NEAT project devoted to asteroid search
(Teegarden et al., 2003). DSS plates are covering the whole
sky while NEAT data exist for the areas within about $40\deg$ from the ecliptic,
except for the Milky Way plane.

\section{Variability search method}

Originally the search for variable objects was carried out in the
vicinity of the 1RXS sources from the bright source catalog only (Voges et
al., 1999) which lists 18806 objects. As of early 2010, we have checked
about 700 northern hemisphere objects without identification in Simbad, NED
and GCVS catalogs, according to the work by Zickgraf et al. (2003). Upon
the discovery of the dwarf nova 1RXS~J231935.0$+$364705 it was noted that the
variability of this star could have been found on the basis of the USNO-B1.0
catalog data alone. Its outburst was detected in the infrared plate taken
on 1995 August 28th, which is why it was included into the USNO-B1.0
catalog with magnitudes: B1$=17.73$, R1$=17.73$,
B2$=18.66$, R2$=17.73$, I$=13.63$. Obviously the color indices $B-R=0.0$,
$R-I=4.1$ do
not correspond to any reasonable spectral energy distribution, even with the
interstellar extinction taken into account and indicate the object's
variability with an amplitude of about 4 magnitudes. 
The fortunate fact has been that the 2nd epoch (POSS-II) Palomar plates of the same
areas of the sky but in different wavelengths, were taken on different dates.
In fact the USNO-B1.0 catalog contains the ``light curves'' in two
passbands, although they consist of only two points, plus one point in the infrared
band. Moreover, for about 60\% of the northern celestial hemisphere in the
areas where nearby plates are overlapping, there are several (2 to 4) images
in each of the three bands $B$, $R$ and $I$. Thus we can detect the object's
variability both by comparing their magnitudes at different epochs (B1 vs.
B2, R1 vs. R2 in the USNO-B1.0 catalog) and by analyzing their $B-R$ and
$R-I$ color
indices, as well as by visual inspection of the photographic plates taken on
different dates in the same wavelengths.

Given that the 1st epoch (POSS-I) Blue and Red Palomar plates were taken during
the same night immediately one after the other, the values of blue and red
magnitudes B1 and R1 can serve as a good estimate of the color index
($\text{B1}-\text{R1}$)
even for the variable stars and non-stellar objects in the USNO-B1.0 catalog.
This is an important criterion which allows us to separate the cold red
($B-R>3$) variable stars of semiregular and Mira types (which can also have
variability with an amplitude up to 2 or more magnitudes), from the systems
of our interest with a blue dwarf or quasi-stellar objects (quasars) who
have blue or white color indices ($B-R \sim 0$). For
example, in the case of 1RXS~J231935.0$+$364705 the value of
$\text{B1}-\text{R1}=0.0$ excludes
the Mira scenario and along with a proper motion of about $14$~mas per year
it allows us to unambiguously identify it as a Galactic cataclysmic variable
star.

In order to search for the variable objects of our interest in the USNO-B1.0
catalog (Monet et al., 2003) we have chosen the following criteria:

\begin{itemize}
  \setlength{\itemsep}{1pt}
  \setlength{\parskip}{0pt}
  \setlength{\parsep}{0pt}
 \item declination $\delta>-10\deg$;
 \item distance from an X-ray source listed in the {\em ROSAT} Bright (Voges et
 al., 1999) or Faint (Voges et al., 2000) catalogs $d<30\arcs$;
 \item $12<\text{R2}<20$;
 \item $\text{B1}-\text{R1}<1.0$;
 \item $|\text{B1}-\text{B2}|>2.0$ or $|\text{R1}-\text{R2}|>2.0$ or
 $|\text{R2}-\text{I}|>2.0$;
 \item ``star/galaxy'' discriminator in USNO-B1.0 is $\geq 7$;
 \item there are no bright stars in the immediate vicinity of the object;
 \item object is not present in the 2MASX catalog of
 extended infrared sources (mostly galaxies, Skrutskie et al., 2006);
 \item object is not present in the NVSS catalog of radio sources (Condon et
 al., 1998);
 \item object is not a known variable star, i.e. it is not listed in the General
Catalogue of Variable Stars (Samus et al., 2007--2009) or the Variable Star
Index (VSX, Watson et al., 2006).
\end{itemize}

The galactic cataclysmic variables and quasars have virtually the same $B-R$
and $R-I$ color indices, as well as very similar X-ray to optical luminosity
ratios. Additional data can help with distinguishing them from each other:
(a) proper motions reported in the USNO-B1.0 catalog.
Obviously quasars should have zero proper motion. If the object has a
reported proper motion of $\sim 10$~mas/year or larger, one can tell with a high level of
confidence that this is a star of our Galaxy. However, quite faint and
distant stars can have proper motions too small to be detected even in $\sim 40$
years having passed between the 1st and 2nd epoch Palomar plates.
(b) 2MASS infrared sky survey data (Skrutskie et al. 2006): 
the infrared $J-K_s$ color index of quasars is usually exceeding $1.0$ unlike 
the stars whose $J-K_s$ lies in the range from
$0.0$ to $1.0$ (see Figure 1 in G{\"a}nsicke et al., 2005). In a number of cases
this allows to make the final choice in favor of one or the other nature of
the object in question. However, additional photometric and/or spectroscopic
observations may be required for the faint sources (generally with $R>18$)
that are absent in 2MASS catalog. We have found two such objects in our
work that do not have a proper motion in USNO-B1.0 and are absent in 2MASS.
We have designated them as objects of unknown nature (QSO?).

The X-ray spectral information available in the 1RXS catalog
in the form of hardness rations HR1 and HR2 (Voges et al., 1999; He et al., 2001)
may provide additional clues about the nature of the sources. For a discussion of typical
X-ray spectral parameters of cataclysmic variables see Verbunt et al. (1997).

\begin{table}[!h]
\setcaptionmargin{0mm}
\onelinecaptionstrue
\captionstyle{flushleft}  
\caption{New variable stars identified with X-ray sources from 1RXS catalog.}
\label{tab:tab1}
\bigskip
\begin{tabular}{ c c r c l l }
  \hline
  \hline
  1RXS & USNO-B1.0 & d\arcs & Range & Constellation~~ & Type \\

  \hline
~~J043445.0$+$030623~~ & ~~0931-0068625~~ & ~~~~$11$ & ~~$17.9$--$20.0 ~R$~~ &
Taurus~~~~ & QSO?~~~~ \\
J072103.3$-$055854 & 0840-0137592 & $27$ & $15.7$--$18.9 ~B$ & Monoceros & CV \\
J164103.6$+$784307 & 1687-0061156 & $19$ & $17.8$--$20.0 ~R$ & Ursa Minor & CV? \\
J174320.1$-$042953 & 0855-0326594 & $ 4$ & $15.1$--$18.1 ~R$ & Ophiuchus & CV \\
J184542.4$+$483134 & 1385-0291789 & $ 3$ & $16.6$--$21.3 ~B$ & Draco & QSO? \\
J184543.6$+$622334 & 1523-0313957 & $13$ & $16.9$--$19.7 ~B$ & Draco & CV \\
J185310.0$+$594509 & 1497-0258111 & $ 4$ & $18.5 ~R$--$15.4 ~I$ & Draco & CV \\
J192926.6$+$202038 & 1103-0421031 & $17$ & $16.6$--$19.0 ~R$ & Vulpecula & CV \\
J194151.4$+$752621 & 1654-0092633 & $ 9$ & $17.8$--$20.5 ~B$ & Draco & CV \\
J222335.6$+$074515 & 0977-0743560 & $22$ & $18.0$--$20.5 ~B$ & Pegasus & CV \\
  \hline
\end{tabular}
\end{table}

\begin{table}[!h]
\setcaptionmargin{0mm}
\onelinecaptionstrue
\captionstyle{flushleft}  
\caption{Magnitudes of the newly identified variable objects in the USNO-B1.0 catalog.}
\label{tab:tab2}
\bigskip
\begin{tabular}{ c c c c c c }
  \hline
  \hline
USNO-B1.0 & B1 & R1 & B2 & R2 & I \\
  \hline
~~0931-0068625~~ & ~~$18.57$~~ & ~~$17.86$~~ & ~~$19.49$~~ & ~~$19.96$~~ & ~~$18.20$~~ \\
0840-0137592 & $15.71$ & $15.17$ & $18.88$ & $16.78$ & $16.80$ \\
1687-0061156 & $18.47$ & $17.82$ & $20.87$ & $19.99$ & ---     \\
0855-0326594 & $16.12$ & $15.15$ & $18.95$ & $18.13$ & $17.10$ \\
1385-0291789 & $21.32$ & $19.81$ & $16.61$ & $19.47$ & $19.04$ \\
1523-0313957 & $16.88$ & $17.28$ & $19.66$ & $19.11$ & $18.52$ \\
1497-0258111 & $17.75$ & $18.66$ & $17.87$ & $18.51$ & $15.44$ \\
1103-0421031 & $17.53$ & $16.60$ & $20.13$ & $19.02$ & $16.62$ \\
1654-0092633 & $17.84$ & $18.25$ & $20.46$ & $19.55$ & $18.22$ \\
0977-0743560 & $17.95$ & $17.72$ & $20.50$ & $17.60$ & $17.63$ \\
  \hline
\end{tabular}
\end{table}

\section{Search results}

Having applied the above criteria to the faint objects of 1RXS catalog we
have obtained a list of 1460 candidates from USNO-B1.0. The analysis of
their distribution over the celestial sphere has shown that over 85\% of
candidates are concentrated in several localized areas and are artifacts of
the erroneous zero-point calibration for one of the catalog's stellar
magnitudes. Besides their proximity in the sky, such objects have
the same color index  (for example,
$\text{B1}-\text{R1}$ or $\text{B2}-\text{R2}$) outside the usual range.
After removing such cases, a list of 203 candidates
remains. Each of them was visually inspected using the digitized Palomar
plates: 6 objects turned out to be ``ghosts'' from the nearby bright stars of
4--8m within $1$--$6\arcm$, 4 were found to be plate defects, 3 were caused by the
absence of 1st epoch Blue plate, 31 are galaxies, 10 are quasars ($J-K_s$ from
$1.3$ to $2.0$), 99 are faint stars likely with the photometry error in one of
filters due to the erroneous zero-point during the digitizing, 31 are pairs
and groups of 3--5 nearby stars, 2 are ``star + galaxy'' pairs, 5 are high
proper motion stars, 1 is a ``white dwarf + red dwarf'' pair (see below), 1 is
a known cataclysmic variable discovered by Catalina Sky Survey on 2009
September 18th and 10 are previously unknown variable objects, 8 of those
are cataclysmic variables and 2 are likely optically violent variable
quasars.

Dwarf novae stay at minimum light (quiescence) for 80 to 99\% of the time 
depending on the variability subtype, hence the probability of detecting
their variability using five stellar magnitudes of the USNO-B1.0 catalog
varies significantly. One can estimate that about 25 to 40\% of UGSU-type
dwarf novae, which are in outburst phase from 5 to 10 per cent of the time,
could be found using this method.

Among bright {\em ROSAT} sources (with {\em ROSAT}/PSPC count rate above $0.05$~cts/s) we have found
two new variable objects -- 1RXS~J184542.4$+$483134 in Draco and
1RXS~J174320.1$-$042953 in Ophiuchus. The former is probably a quasar which
underwent a flare in 1990 during the {\em ROSAT} observations and gradually faded
away over the next years, and the latter -- a cataclysmic variable. Details
for these objects are given in the following section.
The whole list of the newly discovered variables is
presented in Table~\ref{tab:tab1}.

\section{Individual objects}

\begin{figure}[h!]
\setcaptionmargin{5mm}
\onelinecaptionstrue  
\includegraphics[width=0.8\textwidth]{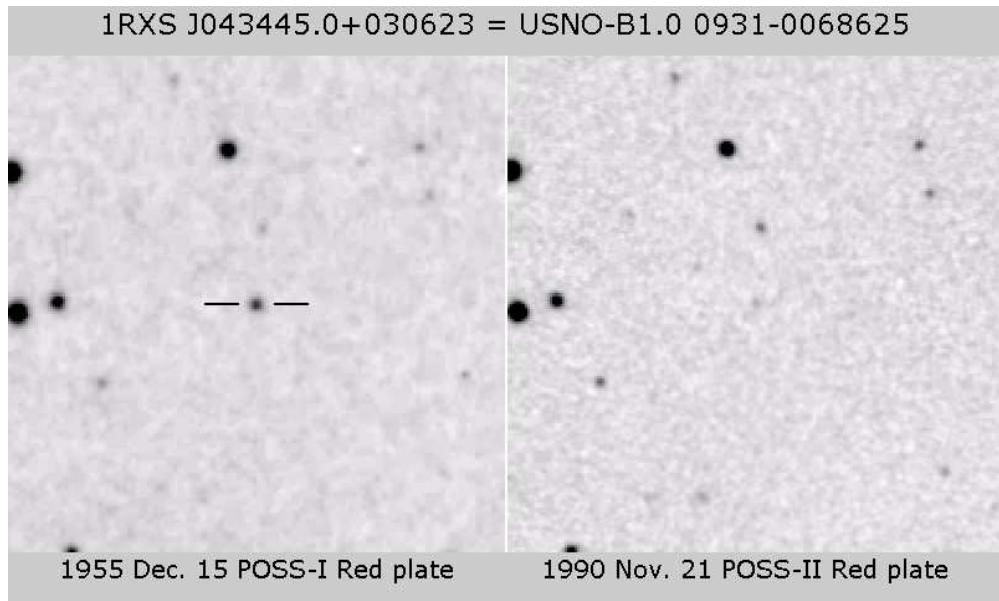}
\captionstyle{normal} \caption{Finder chart of the source
1RXS~J043445.0$+$030623 ($200\arcs \times 200\arcs$).}
\label{fig:J0434p0306}
\end{figure}

{\bf 1RXS~J043445.0$+$030623 = USNO-B1.0~0931-0068625:}
R.A.$=04$:$34$:$44.443$,
Dec.$=+03$:$06$:$16.18$, J2000, proper motion (hereafter $\Delta$R.A., $\Delta$Dec.): ($0$,
$0$)~mas/year. Bai et al. (2007) mention
this object among the Active Galactic Nuclei candidates but with no
spectroscopic or photometric detail. The object is 2.1m brighter on the 1955
Red Palomar 1st epoch plate than on the corresponding 1990 image (see
Fig.~\ref{fig:J0434p0306}). Zero proper
motion in the USNO-B1.0 catalog as well as lack of soft emission in X-ray band
(hardness according to 1RXS catalog is $\text{HR1}=1.00\pm0.43$,
$\text{HR2}=0.83\pm0.53$) favor quasar nature of the object.

\begin{figure}[h!]
\setcaptionmargin{5mm}
\onelinecaptionsfalse 
\includegraphics[width=0.8\textwidth]{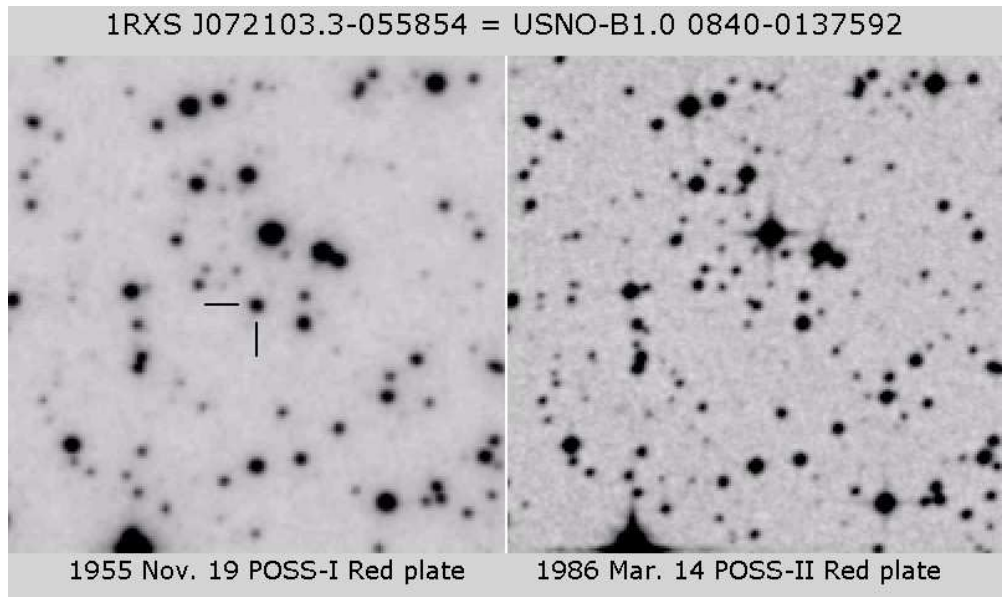}
\captionstyle{normal}
\caption{Finder chart of the source 1RXS~J072103.3$-$055854 ($200\arcs \times 200\arcs$). Note the
faint star to the North-East of the variable at minimum light on the 1986 image.}
\label{fig:J0732m0558}
\end{figure}

{\bf 1RXS~J072103.3$-$055854 = USNO-B1.0 0840-0137592:}
R.A.$=07$:$21$:$02.865$,
Dec.$=-05$:$59$:$20.42$, J2000, proper motion: ($0$, $0$)~mas/year. Despite the zero proper
motion in the USNO-B1.0 catalog, the galactic latitude $b = +3.8$ and the absence of
galaxies and their clusters in the vicinity
(Fig.~\ref{fig:J0732m0558}) favor the galactic
nature of the object.

We have checked 22 photographic plates of the Moscow archive at the
Sternberg Astronomical Institute of Moscow University (area SAO~134264) taken between
November 1986 and February 1994. The star is not visible on 18 of them
(fainter than the plate limit of $17.0$--$17.5~B$), however it has photographic
(blue) magnitude of $16.2$--$16.3$ on two plates of 1986 November 30th,
$16.5$ on the 1992 April 3rd plate and about $17.0$ on 1993 January 19th. The star is
likely a dwarf nova.

\begin{figure}[h!]
\setcaptionmargin{5mm}
\onelinecaptionstrue  
\includegraphics[width=0.8\textwidth]{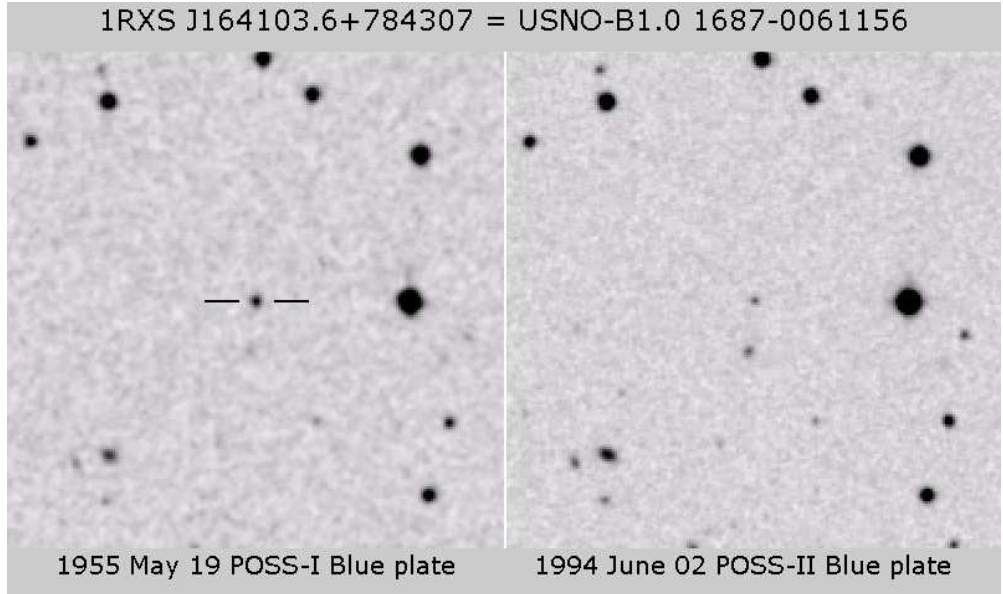}
\captionstyle{normal}
\caption{Finder chart of the source 1RXS~J164103.6$+$784307 ($200\arcs \times 200\arcs$).}
\label{fig:J1641p7834}
\end{figure}

{\bf 1RXS~J164103.6$+$784307 = USNO-B1.0 1687-0061156:}
R.A.$=16$:$40$:$57.850$,
Dec.$=+78$:$42$:$59.26$, J2000, proper motion: ($0$, $0$)~mas/year. There are faint galaxies
present in the vicinity of the object (Fig.~\ref{fig:J1641p7834}). It is possible that the
object is actually a quasar. The hardness in X-ray band according to the {\em ROSAT}
data ($\text{HR1}=0.74\pm0.25$, $\text{HR2}=0.26\pm0.33$) does not allow to make an unambiguous
conclusion about the object's nature. The follow-up observations are
necessary.

\begin{figure}[h!]
\setcaptionmargin{5mm}
\onelinecaptionsfalse 
\includegraphics[width=0.8\textwidth]{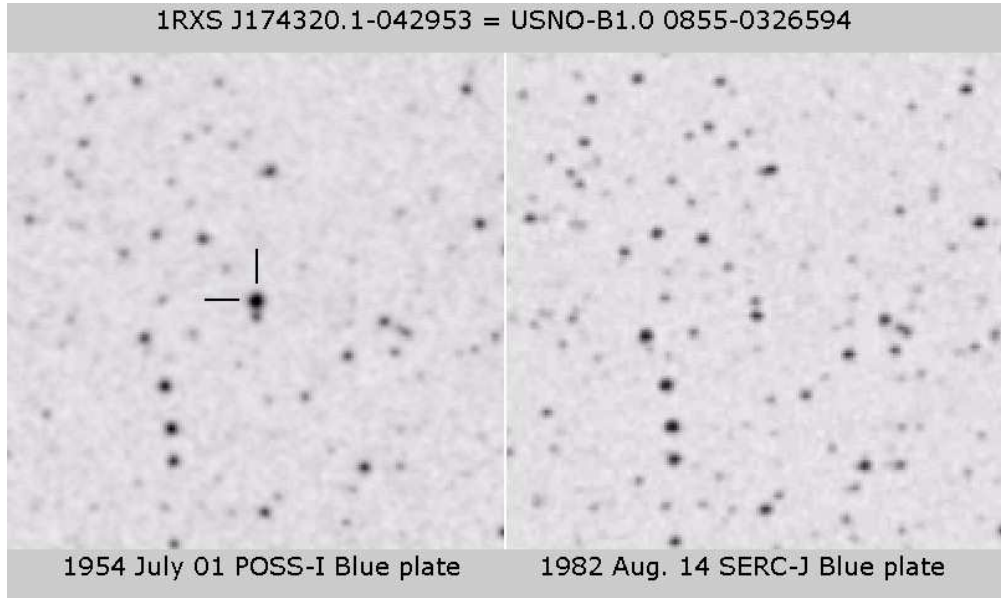}
\captionstyle{normal}
\caption{Finder chart of the source 1RXS~J174320.1$-$042953 ($200\arcs \times
200\arcs$). Note the faint star 6\arcs due South of the variable.}
\label{fig:J1743m0429}
\end{figure}

{\bf 1RXS~J174320.1$-$042953 = USNO-B1.0~0855-0326594:}
R.A.$=17$:$43$:$20.287$,
Dec.$=-04$:$29$:$57.09$, J2000, proper motion: ($18$, $-16$)~mas/year. Star is in outburst on
the Blue and Red POSS-I Palomar plates (1954 July 1st,
Fig.~\ref{fig:J1743m0429}). Color index $J-K_s=0.64$
(according to the 2MASS catalog) and the proper motion clearly indicate that
the object is a cataclysmic variable.

\begin{figure}[h!]
\setcaptionmargin{5mm}
\onelinecaptionsfalse 
\includegraphics[width=0.8\textwidth]{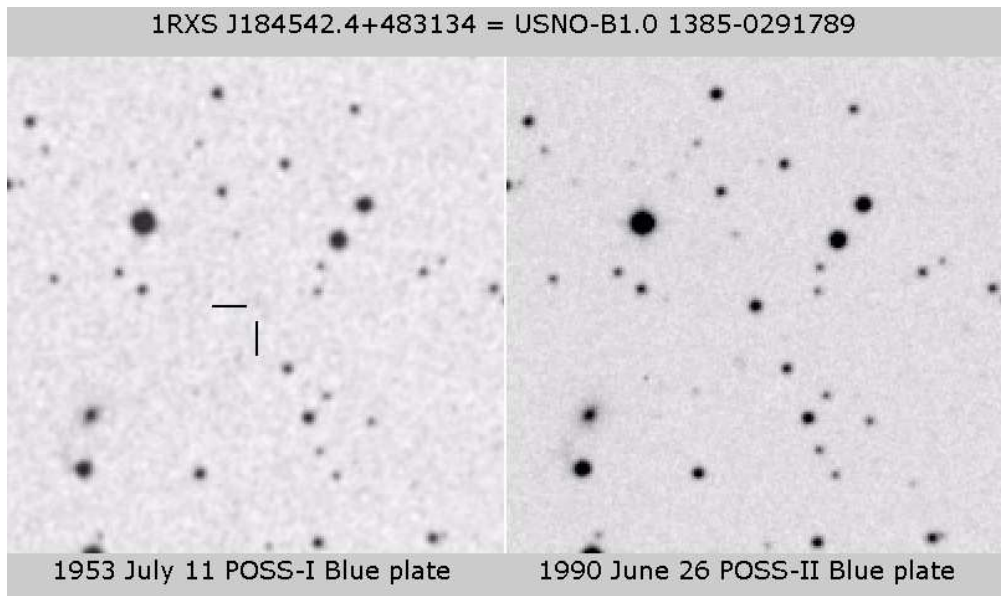}
\captionstyle{normal}
\caption{Finder chart of the source 1RXS~J184542.4$+$483134 ($200\arcs \times
200\arcs$). On the 1953 image (left panel) the object is not visible, its
position being marked with dashes. Note the galaxy $1\farcm4$~South--East (to the bottom left) of the source.}
\label{fig:J1845p4831}
\end{figure}

{\bf 1RXS~J184542.4$+$483134 = USNO-B1.0~1385-0291789:}
R.A.$=18$:$45$:$42.622$,
Dec.$=+48$:$31$:$30.84$, J2000, proper motion: ($0$, $0$)~mas/year. This source is of special
interest. The object was at outburst on the Blue POSS-II Palomar plates
taken on June 26th and July 21st, 1990 then gradually faded during several
years. The source is still clearly seen on the Red plate of 1991 August
7th, while on 1992 August 26th it is already at the limit of sensitivity.
The duration of outburst maximum (no less than 25 days) and monotonous
fading of magnitude during more than two years are hardly consistent with
the behavior of cataclysmic variables.  Most likely the object is a quasar
with large amplitude of variability (at least 5m), and {\em ROSAT} observations
were coincident with its flare. The extragalactic nature of this source is
also supported by the parameters of X-ray hardness in 1RXS catalog
($\text{HR1}=0.03\pm0.15$, $\text{HR2}=0.26\pm0.19$) which are quite typical for quasars.
The finding chart of the object is presented on
Fig.~\ref{fig:J1845p4831}.

\begin{figure}[h!]
\setcaptionmargin{5mm}
\onelinecaptionstrue 
\includegraphics[width=0.8\textwidth]{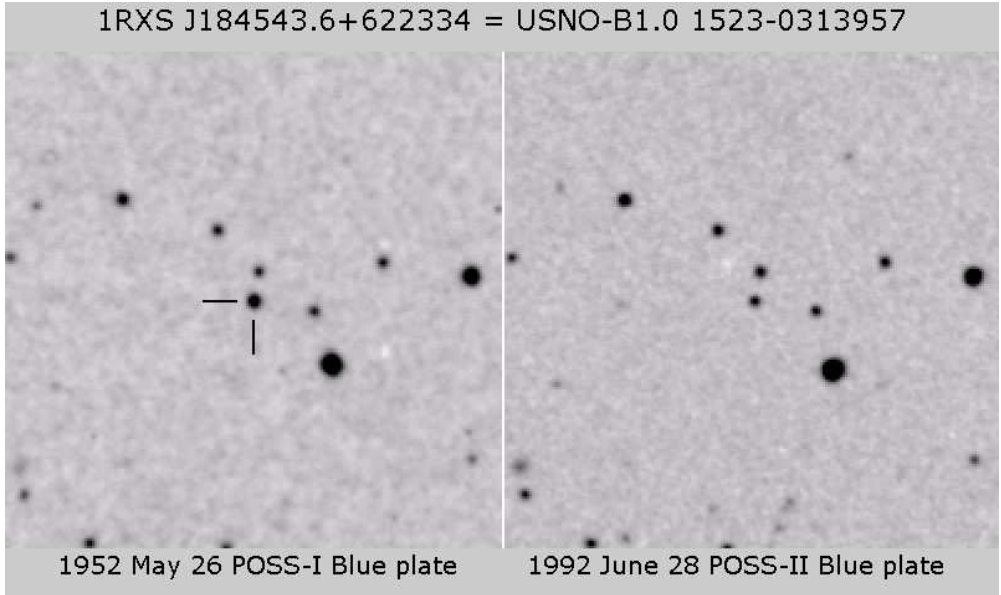}
\captionstyle{normal}
\caption{Finder chart of the source 1RXS~J184543.6$+$622334 ($200\arcs \times 200\arcs$).}
\label{fig:J1845p6223}
\end{figure}

{\bf 1RXS~J184543.6$+$622334 = USNO-B1.0~1523-0313957:}
R.A.$=18$:$45$:$42.285$,
Dec.$=+62$:$23$:$43.99$, J2000, proper motion: ($-2$, $8$)~mas/year. The hardness parameters
in 1RXS catalog ($\text{HR1}=0.52\pm0.23$, $\text{HR2}=0.40\pm0.21$) are corresponding to the
values characteristic of the magnetic cataclysmic variables (for the
polar 1RXS~J005528.0$+$461143 they are equal to $0.37\pm0.19$ and $0.53\pm0.20$,
respectively). The finding chart of the object is presented on
Fig.~\ref{fig:J1845p6223}.

\begin{figure}[h!]
\setcaptionmargin{5mm}
\onelinecaptionsfalse 
\includegraphics[width=0.8\textwidth]{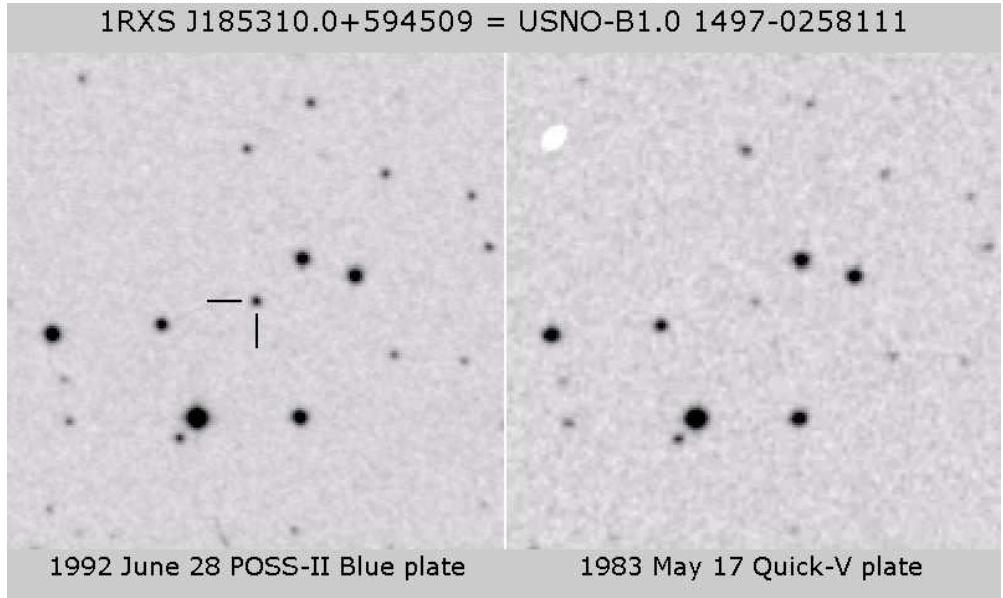}
\captionstyle{normal}
\caption{Finder chart of the source 1RXS~J185310.0$+$594509 ($200\arcs \times
200\arcs$). Left -- Blue POSS-II plate of 1992, right -- photovisual 1983
plate.}
\label{fig:J1853p5945}
\end{figure}

{\bf 1RXS~J185310.0$+$594509 = USNO-B1.0~1497-0258111:}
R.A.$=18$:$53$:$09.556$,
Dec.$=+59$:$45$:$07.22$, J2000, proper motion: ($16$, $12$)~mas/year. Outburst on the infrared
plate: B1$=17.75$, R1$=18.66$, B2$=17.87$, R2$=18.51$, I$=15.44$. X-ray hardness
($\text{HR1}=0.40\pm0.21$, $\text{HR2}=0.29\pm0.25$) also does not contradict the hypothesis of
the magnetic nature of this CV. The finding chart of the object is presented
on Fig.~\ref{fig:J1853p5945}.

\begin{figure}[h!]
\setcaptionmargin{5mm}
\onelinecaptionstrue 
\includegraphics[width=0.8\textwidth]{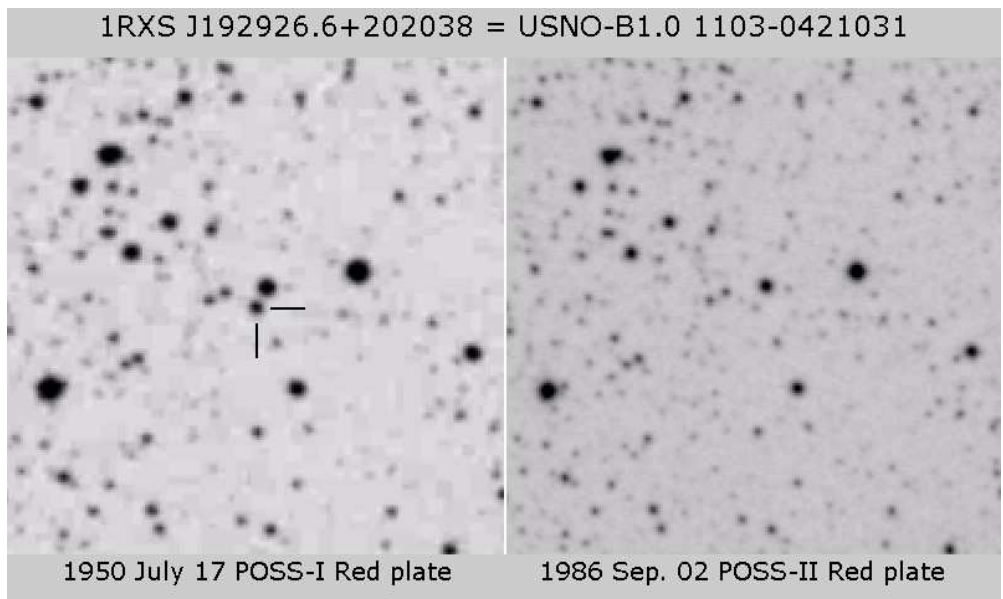}
\captionstyle{normal}
\caption{Finder chart of the source 1RXS~J192926.6$+$202038 ($200\arcs \times 200\arcs$).}
\label{fig:J1929p2020}
\end{figure}

{\bf 1RXS~J192926.6$+$202038 = USNO-B1.0~1103-0421031:}
R.A.$=19$:$29$:$27.820$,
Dec.$=+20$:$20$:$35.36$, J2000, proper motion: ($10$, $10$)~mas/year. Resides in a very dense
star field (galactic latitude $b = +1.2\deg$). X-ray hardness ($\text{HR1}=1.00\pm0.34$, $\text{HR2}=0.22\pm0.39$) is close to similar
parameters for the dwarf nova 1RXS~J231935.0$+$364705 ($1.00\pm0.09$ and
$0.47\pm0.16$, respectively).
The finding chart of the object is presented on
Fig.~\ref{fig:J1929p2020}.

\begin{figure}[h!]
\setcaptionmargin{5mm}
\onelinecaptionstrue
\includegraphics[width=0.8\textwidth]{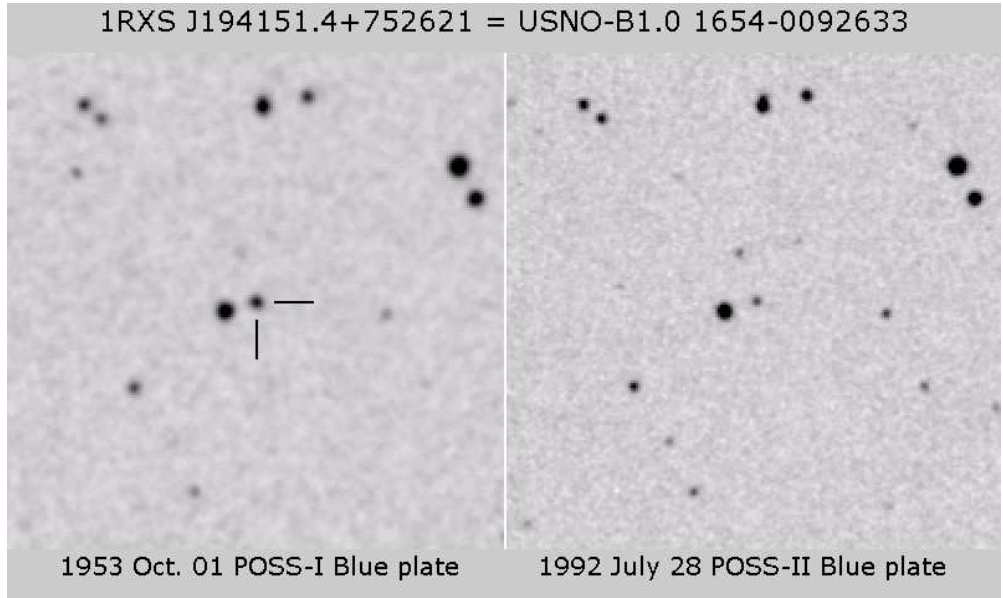}
\captionstyle{normal}
\caption{Finder chart of the source 1RXS~J194151.4$+$752621 ($200\arcs \times 200\arcs$).}
\label{fig:J1941p7526}
\end{figure}

{\bf 1RXS~J194151.4$+$752621 = USNO-B1.0~1654-0092633:}
R.A.$=19$:$41$:$51.597$,
Dec.$=+75$:$26$:$12.81$, J2000, proper motion: ($-6$, $12$)~mas/year. X-ray hardness
($\text{HR1}=0.88\pm0.16$, $\text{HR2}=0.36\pm0.19$) which is typical for dwarf novae along with
rather large amplitude of variability allow us to classify this object
confidently as a cataclysmic variable of U Gem type.
The finding chart of the object is presented on
Fig.~\ref{fig:J1941p7526}.

\begin{figure}[h!]
\setcaptionmargin{5mm}
\onelinecaptionstrue
\includegraphics[width=0.8\textwidth]{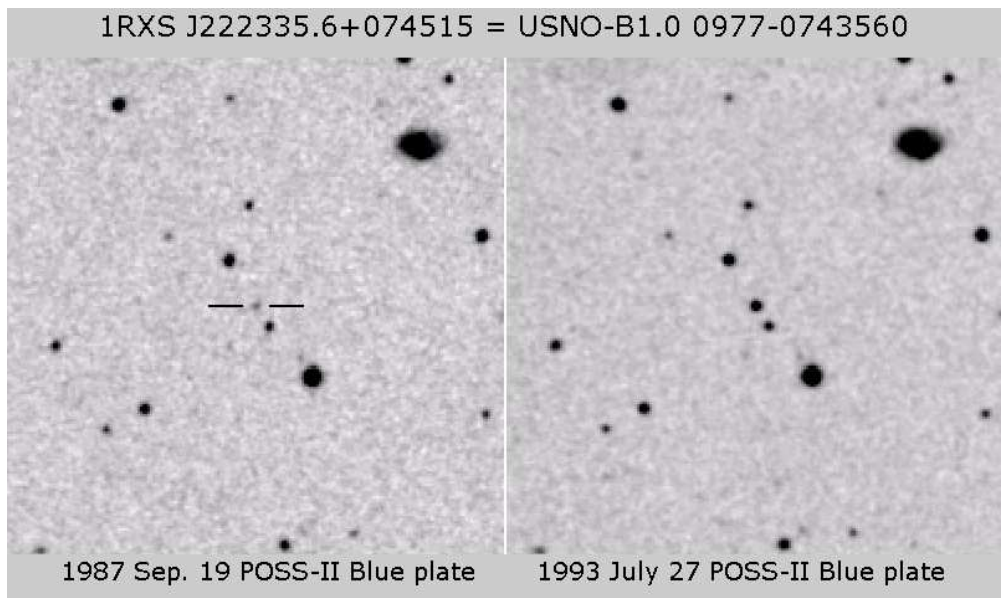}
\captionstyle{normal}
\caption{Finder chart of the source 1RXS~J222335.6$+$074515 ($200\arcs \times 200\arcs$).}
\label{fig:J2223p0745}
\end{figure}

\begin{figure}[h!]
\setcaptionmargin{5mm}
\onelinecaptionstrue 
\includegraphics[width=0.8\textwidth]{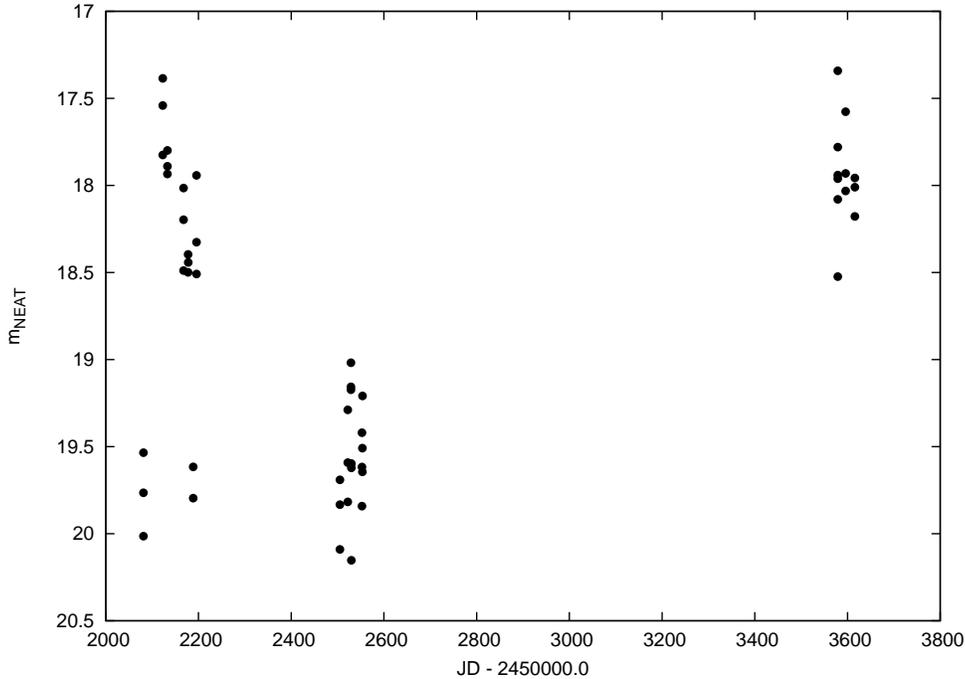}
\captionstyle{normal}
\caption{Light curve of the source 1RXS~J222335.6$+$074515 based on the observations by NEAT.}
\label{fig:RXJ2223.6p0745_NEAT_lightcurve}
\end{figure}

{\bf 1RXS~J222335.6$+$074515 = USNO-B1.0~0977-0743560:}
R.A.$=22$:$23$:$34.159$,
Dec.$=+07$:$45$:$19.27$, J2000, proper motion: ($-12$, $-18$)~mas/year. We have downloaded 52
images of this variable star by the NEAT project from the SkyMorph site
(Pravdo et al., 2005) covering 4 years of observations (June 2001 --
September 2005). The corresponding light curve is shown at
Fig.~\ref{fig:RXJ2223.6p0745_NEAT_lightcurve}. The star
is exhibiting frequent outbursts fading to minimum and rising by more than
2.5m within less than 10~days. This object is likely a dwarf nova of UGSU or
UGZ subtype.
The finding chart of the object is presented on
Fig.~\ref{fig:J2223p0745}.

\section{Red/White dwarf common proper motion pair}

The search for the cataclysmic variables in the USNO-B1.0 catalog has
resulted in an unexpected by-product -- the discovery of the pair
``white dwarf -- red dwarf'' with the common proper motion just below $0.2\arcs$ per
year which could have otherwise remained unnoticed for long. This
binary system is an optical counterpart of the 1RXS~J052436.9$-$053525 X-ray
source. Only one member of the pair with a proper motion is listed in
the UCAC2 (Zacharias et al., 2004) and UCAC3 (Zacharias et al., 2010) catalogs.
Both stars are present in 2MASS catalog but
without a proper motion. The USNO-B1.0 catalog formally includes both stars,
but in fact their data are mixed with each other and the proper motions are given
erroneously. The position of the red dwarf on the 1st
epoch plates (POSS-I) taken in December of 1955 almost coincides with the
position of white dwarf on the 2nd epoch images (red plate of 1994 and
infrared one of 2002). As a result the USNO-B1.0 catalog contains the
record number 0844-0057342 with the following coordinates and peculiar color
indices:
R.A.$=05$:$24$:$36.54$, Dec.$=-05$:$35$:$00.9$, B1$=12.77$, R1$=15.50$, B2$=12.23$,
R2$=14.96$, I$=12.76$.

Visual comparison of 1st and 2nd epoch DSS plates has shown that two stars
with separation about $7\arcs$ at $\text{PA}=159\deg$ are moving together with virtually
identical proper motions about ($-20$, $-160$)~mas/year along the Right
Ascension and Declination respectively, that is the pair's common proper
motion is directed almost along the vector joining the two stars. The data
from the 2MASS catalog for both stars agree with the hypothesis that this pair
is a gravitationally bound system of the white and red dwarfs at the same
distance of about $40$~pc:

2MASS~05243648$-$0535175: $J=12.103\pm0.027$, $H=11.535\pm0.023$, $K_s=11.116\pm0.024$

2MASS~05243627$-$0535103: $J=14.213\pm0.051$, $H=14.260\pm0.058$, $K_s=14.446\pm0.080$

Reid et al. (2007) are providing M$5.0$ spectral class and distance of $37.9$~pc
for the red dwarf but do not mention the second component. 
One can estimate the distance between two components in
the sky plane (lower limit on the size of the binary orbit) as $292$~A.U. With
the mass of the system equal to that of the Sun orbital period would be no
less than $5000$~years. That agrees well with the absence of the detectable
orbital motion of components during $39$ years having passed between 1st and
2nd epoch DSS plates.

\section{Summary}

Using rather simple search method over the publicly available data we have
discovered eight new cataclysmic variables and two Optically Violent
Variable quasar candidates as well as a ``white dwarf -- red dwarf'' common
propper motion pair among the previously unidentified sources of 1RXS
catalog. This result demonstrates the large potential of the Virtual
Observatory concept in identifying peculiar objects. 
Variable objects reported in this paper
are in need of follow-up observations to confirm and
refine their classification. They are all accessible for observatories in
the Northern hemisphere.  The color combined finder charts of the new
variables may be found at \url{http://hea.iki.rssi.ru/~denis/CV-USNO.html}.

\begin{acknowledgments}
Authors are thankful to Dr. N. N. Samus for providing the opportunity to
work with the archival photographic plates of the Moscow collection at SAI
MSU. We are grateful to E.~Angelakis and the anonymous referee for reviewing
the English and Russian versions of this manuscript.
D.D. would like to thank the Program of support of the leading
scientific schools (grant NSh-5069.2010.2). K.S. is a member of the International Max Planck
Research School (IMPRS) for Astronomy and Astrophysics at the
universities of Bonn and Cologne. The work has made an intensive use of
Palomar Observatory Sky Survey plates obtained by the California Institute
of Technology with the funds from NSF, NGS, Sloan Foundation, Samuel Oschin
Foundation and Eastman Kodak Corporation. While preparing the article we
have made use of the SIMBAD database operated at CDS, Strasbourg, France,
and the Variable Star Index (VSX) maintained by the American Association of
Variable Star Observers.
\end{acknowledgments}

\end{document}